\documentclass[seceq]{ptptex}

\usepackage{graphicx}
\usepackage{wrapft}

\usepackage{amsmath}
\usepackage{theorem}
\theorembodyfont{\rmfamily}
\theoremstyle{break}
\newtheorem{Definition}{Definition.}

\title{
Gauge covariant formulation of Wigner representation through deformational quantization 
--Application to Keldysh formalism with electromagnetic field--
}

\author{
Naoyuki \textsc{Sugimoto}$^1$\footnote{E-mail: sugimoto@appi.t.u-tokyo.ac.jp},
Shigeki \textsc{Onoda}$^2$ and
Naoto \textsc{Nagaosa}$^{2,3}$
}

\inst{
$^1$Department of Applied Physics, University of Tokyo, 7-3-1, Hongo, Bunkyo-ku, Tokyo 113-8656\\
$^2$CREST, Department of Applied Physics, University of Tokyo, 7-3-1, Hongo, Bunkyo-ku, Tokyo 113-8656\\
$^3$Correlated Electron Research Center, National Institute of Advanced Industrial Science and Technology, 1-1-1, Higashi, Tsukuba, Ibaraki 305-8562
}

\abst{
We developed a gauge-covariant formulation of the non-equilibrium Green function method for the dynamical and/or non-uniform electromagnetic field by means of the deformational quantization method. 
Such a formulation is realized by replacing the Moyal product in the so-called Wigner space by the star product, and facilitates the order-by-order calculation of a gauge-invariant observable in terms of the electromagnetic field. 
An application of this formalism to the linear response theory is discussed. 
}
\begin{document}
\maketitle

\section{introduction}\label{sec:intro}
The interaction between matter and electromagnetic fields constitutes the most important ingredients of condensed matter physics. 
Especially, since the high intensity fields become available, the non-linear and non-equilibrium responses are of great current interests. 
Therefore a microscopic quantum theory for these processes is called for. 

There already exist established theories for the response to external perturbations. 
The Kubo formalism~\cite{Kubo57} combined with the Matsubara Green function~\cite{Matubara} is the most standard and successful linear response theory. 
However, its extension to higher-order responses to the non-equilibrium field is not straightforward. 
Another method depends on a semiclassical description such as the Boltzmann transport  equation~\cite{Boltzmann1,Boltzmann2}. 
Although appealing to one's intuition, it sometimes leads to different results from what is obtained from the rigorous Kubo formalism. 
In contrast to these methods, the Keldysh formalism~\cite{Keldysh,KadanoffBaym} yields a powerful theoretical framework for directly handling the non-equilibrium Green function in the presence of electromagnetic field. 
Kandanoff and Baym introduced the Wigner distribution function (WDF): $f(X,p)$, where $X^\mu =(T,\mbox{\boldmath $X$})$ and $p^\mu=(\omega ,\mbox{\boldmath $p$})$ are center-of-mass coordinates and relative energy-momenta, respectively, and derived the equation of motion for the WDF. 
The WDF is an appropriate function to describe the interacting, many-particle system, and the equation for its time-evolution is called the Quantum Boltzmann Equation (QBE). 
However, they did not include the electromagnetic field in this formula. 
Altshuler~\cite{Altshuler}, Langreth~\cite{Langreth}, Rammer~\cite{RammerSmith86}, and Mahan~\cite{Mahan,Mahan2} included the electromagnetic field in such formula. 
However, it is not so straightforward to apply these formulas to the non-linear responses. 

Recently, we developed a generic and systematic theoretical framework for the linear and non-linear responses under the external constant electromagnetic field~\cite{ptp1}. 
In that work, we have employed the Wigner representations composed of the set of the center-of-mass time-space coordinates $X$ and gauge covariant momentum  $\pi=p-qA(X)$, where $p$ is the momentum, $q$ is the electric charge and $A(X)$ is the electromagnetic potential. Then, we have accomplished the gauge-covariant formulation of the non-equilibrium Green function in the $(X,\pi)$ space under the constant electromagnetic field, by replacing a conventional product with the Moyal product~\cite{Moyal49}. 
This framework gives a geometric view analogous to the general relativity: the external electromagnetic field is incorporated in the geometry of the Wigner space. 
Corresponding to the commutation relationship $[\hat x^\mu,\hat \pi_\nu]=i\hbar \delta ^\mu_\nu$ between operators, the Wigner space has the noncommutative geometry between time-space coordinate and energy-momentum. 
Furthermore, in the presence of the electromagnetic field, the non-zero commutator $[\hat \pi _\mu,\hat \pi _\nu]=i\hbar qF_{\mu\nu}$ leads also to the noncommutativity within the energy-momentum space. 
It is highly desirable to generalize this geometrical formulation to the dynamical/non-uniform electromagnetic field. 
In this paper, we will show that this generalization is achieved by using a deformational quantization method~\cite{Kontsevich}. 
In the linear order in the electromagnetic field, this formalism completely agrees with the Kubo formalism~\cite{Kubo57}. 

This paper is organized as follows. 
In \S\ref{K}, we begin with a brief review of the Keldysh formulism and the gauge-covariant Dyson equations with external constant electromagnetic fields, which has been developed in the previous paper~\cite{ptp1}. 
For self-containedness, a brief introduction to the star product and deformational quantization~\cite{Kontsevich} is given in \S\ref{s1}. 
\S\ref{s3} constitutes the main body of the present paper, where we present the gauge-covariant formalism taking into account the generic electromagnetic field using the deformational quantization method. 
In \S\ref{Linear}, an application to the linear response theory is given. 

\section{Keldysh formalism}\label{K}
In this section, we briefly review the Wigner space of $(X,\pi)$ and the Moyal product in the case of the constant electromagnetic field, which have been developed in the previous paper~\cite{ptp1}. 

First, we introduce a matrix of Green functions $\underline G$ in the Keldysh space, 
\begin{equation}
\underline{\hat G}=\left(\begin{array}{cc}
\hat G^{R}&2\hat G^{<}\\
0 & \hat G^{A}
\end{array}
\right),
\end{equation}
where the superscripts $R$, $A$, and $<$ denote the retarded, the advanced and the lesser Green functions, respectively. 
They are defined as
\begin{eqnarray}
&&\hat G^R(x_1,x_2)_{\alpha _1,\alpha _2}:=-i\theta (t_1-t_2)\langle [\hat \psi _{\alpha _1}(x_1),\hat \psi ^\dagger _{\alpha _2}(x_2)]_{\mp }\rangle,\\
&&\hat G^A(x_1,x_2)_{\alpha _1,\alpha _2}:=i\theta (t_1-t_2)\langle [\hat \psi _{\alpha _1}(x_1),\hat \psi ^\dagger _{\alpha _2}(x_2)]_{\mp }\rangle,\\
&&\hat G^<(x_1,x_2)_{\alpha _1,\alpha _2}:=\mp i\langle \hat \psi _{\alpha _2}^\dagger (x_2)\hat \psi _{\alpha _1}(x_1)\rangle,
\end{eqnarray}
where $\hat \psi $ is a Bose (upper sign) or a Fermi (lower sign) field, $\hat \psi ^\dagger $ is a Hermitian conjugate of $\hat \psi $, and $x_1=(t_1,\mbox{\boldmath $x$}_1)$, $\alpha _1$ represents an internal degree of freedom. 
Hereafter, we consider electron systems. 
The matrix $\underline {\hat G}$ of the Green functions satisfies the Dyson equations: 
\begin{eqnarray}
\left ((\underline {\hat G}^{(0)-1}-\underline {\hat \Sigma })*\underline {\hat G}\right )_{\alpha _1,\alpha _2}(x_1,x_2) &=&\delta _{\alpha _1,\alpha _2}\delta (x_1-x_2),\label{dyson0}\\
\left(\underline {\hat G}*(\underline {\hat G}^{(0)-1}-\underline {\hat \Sigma })\right )_{\alpha _1,\alpha _2}(x_1,x_2) &=&\delta _{\alpha _1,\alpha _2}\delta (x_1-x_2),\label{dyson}
\end{eqnarray}
where $\underline {\hat G}^{(0)}$ is the unperturbed Green function for free electrons (note that the {\it superscript} ${}^{(0)}$ means the ``unperturbed'' in this paper), 
$\underline{\hat \Sigma }$ is the matrix of the self-energies which arises from electron-electron and/or electron-phonon interactions and potential scattering. 
$(f*g)(x_1,x_2):=\int dx_3 f(x_1,x_3)g(x_3,x_2)$ is the convolution integral. 

Next, we introduce the Wigner representation: the center-of-mass coordinates $X^\mu=(T,\mbox{\boldmath $X$}):=(\frac{1}{2}(t_1+t_2),\frac{1}{2}(\mbox{\boldmath $x$}_1+\mbox{\boldmath $x$}_2))$, and the energy-momenta $p^\mu =(\omega ,\mbox{\boldmath $p$}):=(\omega _1-\omega _2,\mbox{\boldmath $p$}_1-\mbox{\boldmath $p$}_2)$ for relative coordinates~\cite{Mahan,RammerSmith86}. 
Then, the Dyson equations (\ref{dyson0}) and (\ref{dyson}) are rewritten as
\begin{eqnarray}
&&\left ((\underline {\hat G}^{(0)-1}-\underline {\hat \Sigma })\star _{\hbar }\underline {\hat G}\right )(X,p) =1,\label{dy0}\\
&&\left (\underline {\hat G}\star _{\hbar }(\underline {\hat G}^{(0)-1}-\underline {\hat \Sigma })\right )(X,p) =1,\label{dy}
\end{eqnarray}
where the symbol ``$\star _\hbar $'' represents the Moyal product defined by
\begin{eqnarray}
({\hat f}\star _\hbar {\hat g})_{\alpha _1,\alpha _2}(X,p)
&=&\sum _{\alpha _3}\hat f_{\alpha _1,\alpha _3}(X,p)e^{\frac{i\hbar }{2}(\overleftarrow \partial _{X^\mu }\overrightarrow \partial _{p_\mu }-\overleftarrow \partial _{p_\mu }\overrightarrow \partial _{X^\mu })}\hat g_{\alpha _3,\alpha _2}(X,p)\nonumber\\
&\equiv &\sum _{\alpha _3}\hat f_{\alpha _1,\alpha _3}(X,p)\star _\hbar \hat g_{\alpha _3,\alpha _2}(X,p),
\end{eqnarray}
where $\overleftarrow {\partial }$ and $\overrightarrow {\partial }$ represent the derivatives which operate only to the left-hand side and the right-hand side, respectively.  
The average value of the physical observable which is represented by a one-particle operator can be calculated by a one-particle Green function. 
For example, the particle density $\rho(X)$ and the charge current $J_{k}(X)$ are written by $G^<(x)$ as 
\begin{eqnarray}
&&\rho (X)=\frac{\hbar }{i}\int \frac{d^{d+1}p}{(2\pi \hbar ) ^{d+1}}\mathrm {tr}\left [\hat G^<(p,X)\right ],\\
&&J_{k}(X)=q\frac{\hbar }{i}\int \frac{d^{d+1}p}{(2\pi \hbar )^{d+1}}
\mathrm {tr}\bigg[ \left (\nabla _{p_k}\hat H(\mbox{\boldmath $p$}-q\mbox{\boldmath $A$}(X))\right )\hat G^<(p,X)\bigg ].\label{I2}
\end{eqnarray}

Under the uniform and static electromagnetic field $F_{\mu \nu }=\partial _{X^\mu }A_\nu -\partial _{X^\nu }A_\mu ={\rm const.}$, 
we have recently derived the gauge-covariant form of the Dyson equation 
\begin{eqnarray}
&&(\underline {\hat G}^{(0)-1}_0(X,\pi )-\underline {\hat \Sigma }(X,\pi ))\exp (\frac{i\hbar }{2}({\overleftarrow \partial }_{X^\mu }{\overrightarrow \partial }_{\pi _\mu }-{\overleftarrow \partial }_{\pi _\mu }{\overrightarrow \partial }_{X^\mu }+qF^{\mu\nu}\overleftarrow {\partial }_{\pi ^\mu }\overrightarrow {\partial }_{\pi ^\nu }))\underline {\hat G}(X,\pi ) =1,\nonumber\\
&&\\
&&\underline {\hat G}(X,\pi )\exp (\frac{i\hbar }{2}({\overleftarrow \partial }_{X^\mu }{\overrightarrow \partial }_{\pi _\mu }-{\overleftarrow \partial }_{\pi _\mu }{\overrightarrow \partial }_{X^\mu }+qF^{\mu\nu}\overleftarrow {\partial }_{\pi ^\mu }\overrightarrow {\partial }_{\pi ^\nu }))(\underline {\hat G}^{(0)-1}_0(X,\pi )-\underline {\hat \Sigma }(X,\pi )) =1,\nonumber\\
\end{eqnarray}
where we have used the transformation of variables~\cite{Langreth,ptp1}:
\begin{eqnarray}
p_\mu -qA_\mu (X)\mapsto \pi _\mu,\label{v1}\\
f(X,p)\mapsto f(X,\pi ).\label{v2}
\end{eqnarray}
Here $\pi _\mu $ is the mechanical momentum in the Wigner representation with the electric charge $q$ of the particle, and the subscript ${}_0$ represents the zeroth-order terms with respect to $F$. 
All products appearing in the absence of the electromagnetic field is replaced by the Moyal product~\cite{Moyal49}. 
The operator $\exp (\frac{i\hbar }{2}({\overleftarrow \partial }_{X^\mu }{\overrightarrow \partial }_{\pi _\mu }-{\overleftarrow \partial }_{\pi _\mu }{\overrightarrow \partial }_{X^\mu }+qF^{\mu \nu }{\overleftarrow \partial }_{\pi ^\mu }{\overrightarrow \partial }_{\pi ^\nu }))$ has been obtained from $\exp(\frac{i\hbar }{2}(\overleftarrow \partial _{X^\mu }\overrightarrow \partial _{p_\mu }-\overleftarrow \partial _{p^\mu }\overrightarrow \partial _{X^\mu }))$ 
by changing the coordinates from $p^\mu $ to $\pi ^\mu $ and using fact that $F^{\mu \nu }=\mathrm {const}$. 
However, with dynamical/non-uniform electromagnetic field, derivative $\partial _{X^\mu }$ generates $\partial _{X^\kappa }F^{\mu \nu }(\not =0)$ which complicates the expression of the product in the $(X, \pi )$ space. 
Below we will develop a rigorous treatment of these dynamical/non-uniform cases, which enables for example the description of non-linear responses to even oscillating electromagnetic fields. 

\section{Star product and Deformational quantization}\label{s1}
To achieve generalization to the case in general electromagnetic fields, we will use the deformational quantization method~\cite{Kontsevich}. 
It provides the prescription to construct the star product from the usual product $f\cdot g$ and the Poisson bracket $\displaystyle \{f,~g\}:=\sum _{ij}\alpha ^{ij}(\partial _{x^i}f)(\partial _{x^j}g)$. 
Here, $\{f,~g\}$ is assumed to satisfy Jacobi's identity: $\{f,~\{g,~h\}\}+\{g,~\{h,~f\}\}+\{h,~\{f,~g\}\}=0$ and the matrix $\alpha ^{ij}$ is called the Poisson structure, where $f$, $g$ and $h$ are smooth functions on a finite-dimensional manifold $M$. 
Jacobi's identity can be rewritten by the condition for the Poisson structure:
\begin{eqnarray}
\sum _{l=1}^d\left (\alpha ^{il}\partial _{x^l}\alpha ^{jk}+\alpha ^{jl}\partial _{x^l}\alpha ^{kl}+\alpha ^{kl}\partial _{x^l}\alpha ^{ij}\right )=0,
\ \ \ {\rm for}\ \  {}^\forall i,j,k=1\sim d,
\end{eqnarray}
where we have used $\alpha ^{ij}=-\alpha ^{ji}$, and $d$ is the 
dimension of the manifold. 
The star product is written as 
\begin{eqnarray}
&&f\star g:=\sum_{n=0}^\infty C_n(f,g)(i\hbar /2)^n,\label{star}
\end{eqnarray}
where $C_0(f,g):=f\cdot g$ and $C_1(f,g):=\{f,~g\}$. 
Then $C_n(f,g)$ with $n \ge 2$ is defined so as to satisfy the 
associatively property:
\begin{eqnarray}
f\star (g\star h)=(f\star g)\star h.\label{star2}
\end{eqnarray}
For our purpose, the manifold $M$ is the Wigner space. 
Kontsevich~\cite{Kontsevich} found an ingenious method to construct the star product, which we introduce here. 

\vskip 0.5cm
\noindent
\textit{Kontsevich's star product}

Kontsevich invented the diagram technique to calculate the $n$-th order term in $i\hbar /2$ for the expansion of $f\star g$:
\begin{eqnarray}
(f\star g)(x)=f(x)g(x)+\sum^\infty _{n=1}\left(\frac{i\hbar }{2}\right)^n
\sum_{\Gamma \in G_n}w_{\Gamma }B_{\Gamma ,\alpha }(f,g)\label{K*},
\end{eqnarray}
where $\Gamma $, $B_{\Gamma ,\alpha }(f,g)$ and $w_{\Gamma }$ are defined as follows~\cite{Kontsevich}:
\begin{Definition}
$G_n$ is a set of the graphs $\Gamma $ which have $n+2$ vertices and $2n$ edges. 
Vertices are labeled by symbols ``$1$'',$\cdots$,``$n$'',~``$L$'',~``$R$''. 
Edges are labeled by symbols $(k,v)$, where $k=1,\cdots,n$ , $v=1,\cdots,n,L,R$ and $k\not =v$. 
$(k,v)$ means the edge which starts at ``$k$'' and ends at ``$v$''. 
There are two edges starting from each vertex with $k=1,\cdots,n$.
$L$ and $R$ are the exception, i.e., they act only as the end points of the edges.
Hereafter, $V_{\Gamma }$ and $E_{\Gamma }$ represent the set of the vertices and the edges, respectively. 
\end{Definition}
\begin{Definition}
$B_{\Gamma ,\alpha }(f,g)$ is the operator defined as:
\begin{eqnarray}
B_{\Gamma ,\alpha }(f,g)&:=&\sum_{I:E_{\Gamma }\rightarrow \{i_1,\cdots,i_{2n}\}}
\left [\prod ^n_{k=1}\left (\prod _{e\in E_{\Gamma },
e=(k,*)}\partial _{I(e)}\right )\alpha ^{I((k,v_k^1),(k,v_k^2))}\right ]\times \nonumber\\
&&\left[\left (\prod _{e\in E_{\Gamma },e=(*,L)}\partial _{I(e)}\right )f\right]\times 
\left[\left (\prod _{e\in E_{\Gamma },e=(*,R)}\partial _{I(e)}\right )g\right],
\end{eqnarray} 
where, $I$ is a map from the list of edges $((k,v_k^{1,2})),k=1,2,\cdots n$ to integer numbers  $\{i_1,i_2,\cdots,i_{2n}\}$. 
Here, $1\leq i_n\leq d$ and $d$ is a dimension of the manifold $M$. 
$B_{\Gamma ,\alpha }(f,g)$ corresponds to the graph $\Gamma $ in the following way. 
The vertices ``$1$'',$\cdots$,``$n$'', correspond to the Poisson structure $\alpha ^{ij}$. 
$R$ and $L$ correspond to the functions $f$ and $g$, respectively. 
The edge $e=(k,v)$ represents the derivative acting on the vertex $v$. 
Thus the graph $\Gamma $ corresponds to the bidifferetial operator $B_{\Gamma ,\alpha }(f,g)$ by the above rules. 
The simplest diagram for $n=1$ is shown in Fig.~\ref{example} (a), which corresponds to the Poisson bracket $\displaystyle\{f,~g\}=\sum_{i_1,i_2}\alpha ^{i_1i_2}(\partial _{x^{i_1}}f)(\partial _{x^{i_2}}g)$. 
The higher order terms are the generalizations of this Poisson bracket satisfying Eq.~(\ref{star2}). 
Fig.~\ref{example} (b) shown a graph $\Gamma $ with $n=2$ with the list of edges 
\begin{eqnarray}
((1,L),(1,R),(2,R),(2,3)).
\end{eqnarray}
\begin{figure}[t]
\begin{center}
\includegraphics[width=6cm,clip]{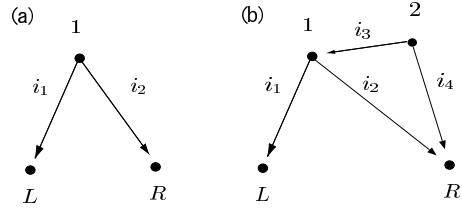}
\end{center}
\caption{(a): The graph $\Gamma \in G_1$ corresponding to Poisson bracket. 
(b): A graph $\Gamma \in G_2$ correspond to the list of edges: $((1,L),(1,R),(2,R),(2,1))\mapsto \{i_1,i_2,i_3,i_4\}$.  
}
\label{example}
\end{figure}
The operator $B_{\Gamma ,\alpha }$ corresponding to this graph is 
\begin{eqnarray}
(f,g)\mapsto \sum _{i_1,\cdots ,i_4}(\partial _{x^{i_3}}\alpha ^{i_1i_2})\alpha ^{i_3i_4}
(\partial _{x^{i_1}}f)(\partial _{x^{i_2}}\partial _{x^{i_4}}g). 
\end{eqnarray}

\end{Definition}
\begin{Definition}
We put the vertices in the upper-half complex plane $H_+:=\{z\in {\mathbf C}|{\rm Im} (z)>0\}$. 
$R$ and $L$ are put at $0$ and $1$, respectively. 
We associate a weight $w_{\Gamma }$ with each graph $\Gamma \in G_n$ as
\begin{eqnarray}
w_{\Gamma }:=\frac{1}{n!(2\pi)^{2n}}\int_{{\cal H}_n}\bigwedge_{k=1}^n\left (d\phi ^h_{(k,v_k^1)}\wedge d\phi ^h_{(k,v_k^2)}\right ),
\end{eqnarray}
where $\phi $ is defined by 
\begin{eqnarray}
\phi ^h_{(k,v)}:=\frac{1}{2i}{\rm Log}\left (\frac{(q-p)(\bar q-p)}{(q-\bar p)(\bar q-\bar p)}\right ).
\end{eqnarray}
$p$ and $q$ are the coordinates of the vertices ``$k$'' and ``$v$'', respectively. 
$\bar p$ is complex conjugate of $p$. 
${\cal H}_n$ represents the space of configurations of $n$ numbered pair-wise distinct points on $H_+$: 
\begin{eqnarray}
{\cal H}_n:=\{ (p_1,\cdots ,p_n)|p_k\in H_+,~p_k\not =p_l~\text{for}~k\not =l\}.
\end{eqnarray}
Here we assume that $H_+$ has the Poincare metric: $ds^2=(d({\rm Re}(p))^2+d({\rm Im}(p))^2)/({\rm Im}(p))^2$, $p\in H_+$. 
From the metric, we can calculate a geodesic. 
Note that, $\phi ^h(p,q)$ is the angle with the geodesics which is defined by $(p,q)$ and $(\infty ,p)$. 
Namely, $\phi ^h(p,q)=\angle pq\infty $. 
For example, $w_\Gamma$ corresponding to the Fig.~\ref{example} (a) is
\begin{equation}
w_1=\frac{2}{1!(2\pi )^2}\int _{{\cal H}_1}
d\frac{1}{2i}{\rm Log}\left (
\frac{p^2}{\overline p^2}\right )\wedge
d\frac{1}{2i}{\rm Log}\left (
\frac{(1-p)^2}{(1-\overline p)^2}\right ) =1,\label{w1}
\end{equation}
where we have included the factor ``$2$'' arisen from the interchange between two edges in $\Gamma $. 
$w_{\Gamma }$ corresponding to the Fig.~\ref{example} (b) is
\begin{eqnarray}
w_{\Gamma_{(b)}}&=&\frac{1}{2!(2\pi )^4}\int _{{\cal H}_2}
d\frac{1}{i}{\rm Log}\left (\frac{p_1}{\overline p_1}\right )\wedge
d\frac{1}{i}{\rm Log}\left (\frac{1-p_1}{1-\overline p_1}\right )\wedge
d\frac{1}{i}{\rm Log}\left (\frac{p_2}{\overline p_2}\right )\wedge 
d\frac{1}{2i}{\rm Log}\left (\frac{(p_1-p_2)(\overline p_1-p_2)}{(p_1-\overline p_2)(\overline p_1-\overline p_2)}\right )
\nonumber\\
&=&\frac{1}{2!(2\pi )^4}\int _{{\cal H}_2}
d\frac{1}{i}{\rm Log}\left (\frac{p_1}{\overline p_1}\right )\wedge
d\frac{1}{i}{\rm Log}\left (\frac{1-p_1}{1-\overline p_1}\right )\nonumber\\
&&\wedge
d\left (2\arg (p_2)\right )\wedge 
d|p_2|\frac{\partial }{\partial |p_2|}\frac{1}{2i}{\rm Log}\left (\frac{(p_1-p_2)(\overline p_1-p_2)}{(p_1-\overline p_2)(\overline p_1-\overline p_2)}\right )
\nonumber\\
&=&\frac{1}{2!(2\pi )^4}\int _{{\cal H}_2}
d\frac{1}{i}{\rm Log}\left (\frac{p_1}{\overline p_1}\right )\wedge
d\frac{1}{i}{\rm Log}\left (\frac{1-p_1}{1-\overline p_1}\right )\wedge
d\frac{1}{i}{\rm Log}\left (\frac{p_2}{\overline p_2}\right )\wedge 
d\frac{1}{i}{\rm Log}\left (\frac{1-p_2}{1-\overline p_2}\right )\nonumber\\
&=&\frac{w_1^2}{2!}\nonumber\\
&=&\frac{1}{2},\label{w2}
\end{eqnarray}
where $p_1$ and $p_2$ are the coordinates of vertices ``$1$'' and ``$2$'', respectively. 
Here, we have used the following facts:
\begin{eqnarray}
\int _0^\infty d|p_2|\partial _{|p_2|}{\rm Log}\left (
\frac{(p_1-p_2)(\overline p_1-p_2)}{(\overline p_1-p_2)(\overline p_1-\overline p_2)}
\right )&=&\lim _{\Lambda \to \infty }{\rm Log}\left (
\frac{(p_1-\Lambda e^{i\arg (p_2)})(\overline p_1-\Lambda e^{i\arg (p_2)})}{(p_1-\Lambda e^{-i\arg (p_2)})(\overline p_1-\Lambda e^{-i\arg (p_2)})}\right )\nonumber\\
&=&\lim _{\Lambda \to \infty }{\rm Log}\left (
\frac{(1-\Lambda e^{i\arg (p_2)})(1-\Lambda e^{i\arg (p_2)})}{(1-\Lambda e^{-i\arg (p_2)})(1-\Lambda e^{-i\arg (p_2)})}\right ),\nonumber\\
\int _{|p_1|\gtrsim \Lambda }
d{\rm Log}\left (\frac{p_1}{\overline p_1}\right )\wedge
d{\rm Log}\left (\frac{1-p_1}{1-\overline p_1}\right )
&\stackrel{\Lambda \rightarrow \infty }{\longrightarrow }&
\int _{|p_1|\gtrsim \Lambda }d{\rm Log}\left (\frac{p_1}{\overline p_1}\right )\wedge 
d{\rm Log}\left (\frac{p_1}{\overline p_1}\right )=0.
\end{eqnarray}

Generally speaking, the integrals are entangled for $n \geq 3$ graphs, and the weight of these are not so easy to evaluate as Eq.~(\ref{w2}). 
\end{Definition}

\section{Gauge-covariant formulation with electromagnetic fields}\label{s3}
\begin{figure}[t]
\begin{center}
\includegraphics[width=8cm,clip]{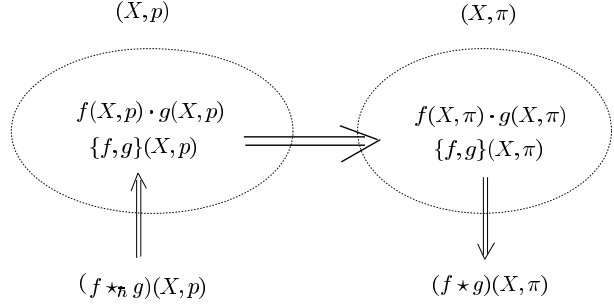}
\end{center}
\caption{This figure represents the derivation of the LHS of the gauge-covariant Dyson equation, where $f$ and $g$ are certain functions in the Dyson equation. 
The derivation has three steps. 
First, we obtain the Poisson structure of the Dyson equation with  Wigner space $(X,p)$.
This Poisson structure is obtained by using the gradient expansion~\cite{RammerSmith86}, which is represented by the symbol ``$\Uparrow $''. 
Secondly, we perform the variable transformation, 
$p_\mu -qA_\mu (X)\mapsto \pi _\mu$, which is represented by the symbol ``$\Longrightarrow $''. 
Thirdly, explicitly gauge-covariant Dyson equations are given by using 
the deformational quantization method which is represented by the symbol ``$\Downarrow $''. }
\label{Dtab}
\end{figure}

In this section, we derive the gauge-covariant Dyson equation. 
Since we can use the deformational quantization method, we only need $C_0$ and $C_1$ in (\ref{star}). 
We should calculate the zeroth-order and the first-order terms which are given by expanding the Dyson equation (\ref{dyson}) with respect to $\hbar $ and by the variable transformation $p_\mu -qA_\mu (X)\mapsto \pi _\mu$. 
The derivation is organized by three steps as shown in Fig \ref{Dtab}.

\subsection{Transformation of variables and Poisson structure}
\vskip 0.5cm
\noindent
Step. 1: We derive the zeroth-order and the first-order terms of the Dyson equations with respect to $\hbar $ by the gradient expansion. 
Expanding the left hand side (LHS) of Eq.~(\ref{dyson0}) with respect to $\hbar$, we obtain the first two terms as 
\begin{eqnarray}
&&((\underline {\hat G}^{(0)-1}_0(x)-\underline {\hat \Sigma }(x))\cdot \underline {\hat G}(x)),\label{0order}\\
&&\frac{i\hbar }{2}\left\{(\underline {\hat G}^{(0)-1}_0(x)
-\underline {\hat \Sigma }(x)),\underline {\hat G}(x))\right\},\label{1order}
\end{eqnarray}
where $x:=(X,p)$ and the dimension of the Wigner space is $4+4$. 
$\displaystyle\{f,g\}:=\sum_{i,j}\alpha_0^{ij}(\partial_{x^i} f)
(\partial _{x^j}g)$ is the Poisson bracket with the Poisson structure 
$\alpha_0^{ij}$ $(i,j=1\sim 4+4)$ given by: 
\begin{eqnarray}
&&\alpha _0^{ij}=\left(
\begin{array}{cc}
0&\eta ^{\mu \nu }\\
-\eta ^{\mu \nu }&0
\end{array}
\right),
\end{eqnarray}
and $\eta ^{\mu \nu}$ is the Minkowski metric 
$\eta =\mathrm{diag}\{-1,1,1,1\}$. 

\vskip 0.5cm
\noindent
Step. 2: Next, we introduce the transformation of variable (\ref{v1}) and (\ref{v2}). 
This step corresponds to the calculation of $C_0$ and $C_1$ in definition (\ref{star}) in $(X,\pi )$ space. 
Using the transformations (\ref{v1}) and (\ref{v2}), we can write the derivative of $f$ and $g$ as follows:
\begin{eqnarray}
&&(\partial _{X^\mu }f(X,p))(\partial _{p_\mu }g(X,p))-(\partial _{p_\mu }f(X,p))(\partial _{X^\mu }g(X,p))\nonumber\\
&=&
(\partial _{X^\mu }f(X,\pi ))(\partial _{\pi _\mu }{g}(X,\pi ))-
(\partial _{\pi _\mu }{f}(X,\pi ))(\partial _{X^\mu}{g}(X,\pi ))\nonumber\\
&&
-q\left (
(\partial _{X^\mu}A_\nu (X))(\partial _{\pi _\nu }{f}(X,\pi ))(\partial _{\pi _\mu }{g}(X,\pi ))
-(\partial _{X^\nu}A_\mu (X))(\partial _{\pi \mu }{f}(X,\pi ))(\partial _{\pi _\nu }{g}(X,\pi ))\right )\nonumber\\
&=&
(\partial _{X^\mu }{f}(X,\pi ))(\partial _{\pi _\mu }{g}(X,\pi ))-(\partial _{\pi _\mu }{f}(X,\pi ))(\partial _{X^\mu}{g}(X,\pi ))
+qF_{\mu \nu }(\partial _{\pi _\mu }{f}(X,\pi ))(\partial _{\pi _\nu }{g}(X,\pi )),
\nonumber\\\label{R}
\end{eqnarray}
where $F$ is the field strength: $F_{\mu\nu}=\partial_{X^\mu} A_\nu-\partial_{X^\nu}A_{\mu}$. 
Accordingly the Poisson structure is transformed as:
\begin{eqnarray}
\alpha _0^{ij}\mapsto \alpha ^{ij}&:=&
\left(
\begin{array}{cc}
0&\eta ^{\mu \nu }\\
-\eta ^{\mu \nu }&qF^{\mu \nu }(X)
\end{array}
\right )\nonumber\\
&=&\left(
\begin{array}{cc}
0&\eta ^{\mu\nu}\\
-\eta ^{\mu\nu}&0
\end{array}
\right)
+\left(
\begin{array}{cc}
0&0\\
0&qF^{\mu\nu}
\end{array}
\right )\nonumber\\
&\equiv &\alpha ^{ij}_0+\alpha ^{ij}_F.\label{P}
\end{eqnarray}
By this replacements, 
the zeroth-order terms and the first-order terms are transformed as 
\begin{eqnarray}
&&((\underline {\hat G}^{(0)-1}_0(\tilde x)-\hat {\underline \Sigma }(\tilde x))\cdot \hat {\underline G}(\tilde x)),\label{eq:gauge-invDyson0}\\
&&\frac{i\hbar }{2}\left\{(\underline {\hat G}^{(0)-1}_0(\tilde x)-\hat {\underline \Sigma }(\tilde x)),
~\hat {\underline G}(\tilde x))\right\},\label{eq:gauge-invDyson1}
\end{eqnarray}
where $\tilde x:=(X,\pi )$. 
In (\ref{eq:gauge-invDyson1}), $\displaystyle\{f,g\}:=\sum_{i,j}\alpha ^{ij}(\partial _{\tilde x^i}f)(\partial _{\tilde x^j}g)$ is also the Poisson bracket, i.e., satisfies Jacobi's identity. 
Hereafter, the tilde in $\tilde x$ will be omitted for simplicity. 

\subsection{Gauge-covariant formulation}
\vskip 0.5cm
\noindent
Step 3: From the 0th-order and the first-order terms obtained in the previous section, gauge-covariant-Dyson equations are given by using the deformational quantization method. 
By constructing the star product from (\ref{eq:gauge-invDyson0}) 
and (\ref{eq:gauge-invDyson1}), the Dyson equation reads: 
\begin{eqnarray}
((\underline {\hat G}^{(0)-1}_0(X,\pi)-\hat {\underline \Sigma}(X,\pi ))\star \hat {\underline G}(X,\pi )=1,\label{gauge-invDyson1}
\end{eqnarray}
where the symbol ``$\star $'' is star product which is given the Poisson structure $\alpha $ with the  Wigner space $(X,\pi )$. 

Similarly, 
\begin{eqnarray}
\hat {\underline G}(X,\pi )\star ((\underline {\hat G}^{(0)-1}_0(X,\pi )-\hat {\underline \Sigma }(X,\pi ))=1.\label{gauge-invDyson2}
\end{eqnarray}

Note that the difference between Eqs.~(\ref{dy0}) and (\ref{gauge-invDyson1}) is only in the definition of the product. 
Namely, to obtain the explicitly gauge-covariant Dyson equation, we only change the product ``$\star _\hbar$'' to ``$\star $''. 

\vskip 0.5cm
\noindent
\textit {Response to electromagnetic field}

Let's recall the diagram rules for Kontsevich's star product. 
The rules relates a bidifferential operator to a graph $\Gamma \in G_n$. 
$G_n$ is a set of $n$th-order graphs with respect to $\hbar $, and each graph $\Gamma \in G_n$ is constructed by $n+2$ vertices which are labeled as ``$1$'', ``$2$'', $\cdots $, ``$n$'', ``$L$'' and ``$R$'', and by $2n$ edges which start from the vertex ``$k$'', $k\not =L$ or $R$.  
The vertices which are labeled as ``$1$'', ``$2$'', $\cdots $, ``$n$'', correspond to $\alpha $. 
The vertices which are labeled as ``$L$'' and ``$R$'' correspond to $\underline {\hat G}$ or $(\underline {\hat G}^{(0)-1}_0-\underline {\hat \Sigma })$. 
We separate the Poisson structure into two parts $\alpha _0+\alpha _F$ as (\ref{P}) and expand $\underline {\hat G}$ and $(\underline {\hat G}^{(0)-1}_0-\underline {\hat \Sigma })$ in terms of $F$: 
\begin{eqnarray}
\underline {\hat G}&=&\underline {\hat G}_0+\underline {\hat G}_F+\underline {\hat G}_{F^2}+\cdots \nonumber\\
(\underline {\hat G}^{(0)-1}_0-\underline {\hat \Sigma })&=&(\underline {\hat G}^{(0)-1}_0-\underline {\hat \Sigma }_0)-\underline {\hat \Sigma }_F-\underline {\hat \Sigma }_{F^2}-\cdots ,
\end{eqnarray}
where subscript ${}_0$ and ${}_{F^k}$ represent zeroth order and $k$-th order terms with respect to $F$, respectively. 
Thus the contributions corresponding to each graph are further classified according to the order in $F$. 
Now, the vertices ``$1$'', ``$2$'', $\cdots $, ``$n$'' correspond to $\alpha _0$ or $\alpha _F$. 
The numbers of $\alpha _F$ and $\alpha _0$ are denoted by $n_{\alpha _F}$ and $n_{\alpha _0}$, the vertices ``$L$'' and ``$R$'' correspond to $f_{F^l}$ and $g_{F^m}$, receptivity, where $f$ and $g$ represent $\underline {\hat G}$ or $(\underline {\hat G}^{(0)-1}_0-\underline {\hat \Sigma })$, and $l,m$ are some nonnegative integers. 
The order of the graph with respect to $F$ is given by $l+m+n_{\alpha _F}$. 
The Kontsevich's diagram rules applied to the present case read as
\begin{enumerate}
\item $n=n_{\alpha _F}+n_{\alpha _0}$ vertices ``$1$'', ``$2$'', $\cdots $, ``$n$'' are put in the upper-half complex plan 
$H_+:=\{z\in {\mathbf C}|{\rm Im}(z)> 0\}$ and two vertices 
``$L$'' and ``$R$'' are put at $0$ and $1$, respectively. 
\item Vertices ``$1$'', ``$2$'', $\cdots $, ``$n$'' have two edges $(k,v_k^{1,2}), k=1\sim n, v_k^{1,2}=1\sim n,L,R$ and $k\not =v_k^1, k\not =v_k^2$, where each edge $(k,v)$ starts at vertex ``$k$'' and ends at any other vertices ``$v$''. 
\item The set of vertices and edges are written by $\Gamma $. 
$G_n$ is defined by the set of all graph of $n+2$ vertices and $2n$ edges. 
$G_n\in \Gamma $.  
\item Vertices ``$1$'', ``$2$'', $\cdots $, ``$n_{\alpha _F}$'' correspond to $\alpha _F$, vertices ``$n_{\alpha _F}+1$'', ``$n_{\alpha _F}+2$'', $\cdots $, ``$n_{\alpha _F}+n_{\alpha _0}=n$'' correspond to $\alpha _0$, and ``$L$'' and ``$R$'' correspond to $f_{F^l}$ and $g_{F^m}$, respectively. 
\item The operator $B_{\Gamma ,\alpha }$ is constructed as follows. 
Edge $(k,v)$ corresponds to the differentiation with respect to $x$ acting on a function corresponding to vertex ``$v$'' $(l=1\sim n,l\not =k)$, ``$L$'' or ``$R$''. 
Thus a set of the vertex ``$k$'' and the two edges $(k,v_k^{1,2})$ corresponds to bidifferential operator $\sum \limits _{ij}\alpha _{0,{\rm or},F}^{ij}\partial _{x^i}(h_{v_k^1})\partial _{x^j}(h_{v_k^2})$, where $h_{v}$ is the function corresponding to the vertex $v$. 
\item The weight $w_\Gamma $ of $\Gamma $ is defined by 
\begin{eqnarray}
&&w_{\Gamma }:=\frac{1}{n!(2\pi )^n}\int _{{\cal H}_n}\bigwedge _{k=1}^n\left (d\phi ^h_{(k,v_k^1)}\wedge d\phi ^h_{(k,v_k^2)}\right ),\nonumber\\
&&\phi ^h_{(k,v)}=\frac{1}{2i}{\rm Log}\left (
\frac{(q-p)(\overline q-p)}{(q-\overline p)(\overline q-\overline p)}
\right ),\nonumber
\end{eqnarray}
where $p$ and $q$ are coordinates of vertices ``$k$'' and ``$v$'' in $H_+$, respectively, and ${\cal H}_n$ is the space of configurations on $n$ numbered pair-wise distinct points on $H_+$: ${\cal H}_n=\{(p_1,\cdots ,p_n)|p_k\in H_+,p_k\not =p_l~{\rm for}~k\not =l\}$. 
\item The $r$th-order star product in terms of $F$ is given by
\begin{eqnarray}
(f_{F^l}\star g_{F^m})_{F^{r}}=\sum \limits_{n_{\alpha _0}=0}^\infty (i\hbar /2)^{n_{\alpha _F}+n_{\alpha _0}}\sum _{\Gamma \in G_{n_{\alpha _F}+n_{\alpha _0}}}w_\Gamma B_{\Gamma ,\alpha }(f_{F^l},g_{F^m}),\nonumber
\end{eqnarray}
where $r=l+m+n_{\alpha _F}$. 
\end{enumerate}

\vskip 0.5cm
\noindent
Additional diagram rules:
\begin{list}{}{}
\item[A1.] Two edges of each vertex corresponding to $\alpha _F$ connect with both vertices ``$L$'' and ``$R$''. 
\item[A2.] Edges of the vertices corresponding to $\alpha _0$ do not connect with the vertices corresponding to $\alpha _0$, because $\eta $ is the constant metric. 
\item[A3.] At least one edge of the vertex corresponding to $\alpha _0$ connects with ``$L$'' or ``$R$''. \\
\end{list}

To calculate $(f\star g)_{F^r}|_{\genfrac{}{}{0pt}{}{f=f_{F^l}}{g=g_{F^m}}}$, first we separate the graph $\Gamma $ into $\Gamma _{\alpha _F}$ and $\Gamma _{\alpha _0}$. 
$\Gamma _{\alpha _F}$ is the graph consisted by vertices corresponding to $\alpha _F$, and ``$L$'' and ``$R$'', and edges starting from these vertices. 
$\Gamma _{\alpha _0}$ is the rest of the graph 
$\Gamma $ without $\Gamma _{\alpha _F}$. 
Namely, $\Gamma _{\alpha _0}$ is the set of the vertices corresponding to $\alpha _0$ which are labeled by ``$k$'' ($k=n_{\alpha _F}\sim (n_{\alpha _F}+n_{\alpha _0})$) and edges $(k,v_k^{1,2})$. 
Next, we calculate the weight $w_{n_{\alpha _F}}$ and the operator $B_{\Gamma _{\alpha _F},\alpha _F}$ corresponding to $\Gamma _{\alpha _F}$, and later those for $\Gamma _{\alpha _0}$. 

\vskip 0.5cm
\noindent
\textit{Separation of graph $\Gamma $}

We now prove $w_{\Gamma }B_{\Gamma ,\alpha }=w_{n _{\alpha _0},n _{\alpha _F}}B_{\Gamma _{\alpha _0},\alpha _0}\cdot w_{n _{\alpha _F}}B_{\Gamma _{\alpha _F},\alpha _F}$, where $w_{n _{\alpha _0},n _{\alpha _F}}=\frac{n_{\alpha _F}!}{(n_{\alpha _F}+n_{\alpha _0})!}\cdot \frac{w_1^{n_{\alpha _0}}}{n_{\alpha _0}!}$, $w_{n _F}=\frac{w_1^{n_{\alpha _F}}}{n_{\alpha _F}!}$ and $w_1$ is given by Eq.~(\ref{w1}). 
\begin{figure}[t]
\begin{center}
\includegraphics[width=3cm,clip]{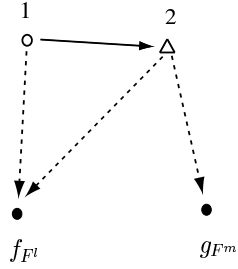}
\end{center}
\caption{A four vertices graph, where the white circle and the white triangle represent $\alpha _0$ and $\alpha _F$, respectively, and the dotted arrow and the real arrow represent the derivative with respect to $\pi $ and $X$, respectively. 
}
\label{graph0.eps}
\end{figure}

First we consider the case where the edges of the vertices corresponding to $\alpha _0$ do not connect with $\alpha _F$. 
In this case, 
$w_{n _{\alpha _0},n _{\alpha _F}}=
\frac{n_{\alpha _0}!n_{\alpha _F}!}{(n_{\alpha _F}+n_{\alpha _0})!}\cdot 
\frac{w_1^{n_{\alpha _0}}}{n_{\alpha _0}!}$. 
Secondly we consider the graph which consists of four vertices corresponding to $\alpha _0$, $\alpha _F$, and ``$L$''and ``$R$'' as shown in Fig.~\ref{graph0.eps}. 
We also assume that one edge of the vertex corresponding to $\alpha _0$ connect with vertex corresponding to $\alpha _F$. 
In this case, from additional diagram rule A3, another edge of the vertex has to connect with ``$L$'' or ``$R$''. 
Since we can exchange the role ``$R$'' and ``$L$'' by the variable transformation $p\mapsto 1-p$, ($p\in H_+$), we can assume that one edge of the vertex corresponding to $\alpha _0$ connect with ``$L$''. 
The weight $w_{\Gamma }$ in this case is given by Eq.~(\ref{w2}), i.e., the integrals for the weight is given by replacing coordinate of the vertex corresponding to $\alpha _F$ by coordinate of ``$R$'' (or ``$L$'') in $H_+$. 
From additional diagram rules, above result of integrals is applied to every graph. 
Namely, the position of each vertex corresponding to $\alpha _0$ and $\alpha _F$ can be move independently in integrals, and the entangled integral does not appear. 
Therefore the weight $w_{n_{\alpha _0},n_{\alpha _F}}$ of a graph $\Gamma _{\alpha _0}$ only depend on the number of vertices corresponding to $\alpha _0$ and $\alpha _F$. and $w_\Gamma =w_{n _{\alpha _0},n_{\alpha_F}} \cdot w_{n _{\alpha _F}}$ holds generally. 
Finally, we can obtain $w_{n _{\alpha _0}}=\frac{n_{\alpha _F}!n_{\alpha _0}!}{(n_{\alpha _F}+n_{\alpha _0})!}\frac{w_1^{n_{\alpha _0}}}{n_{\alpha _0}!}$ and separate the operator $w_{\Gamma }B_{\Gamma ,\alpha }$ into $w_{n _{\alpha _0}}B_{\Gamma _{\alpha _0},\alpha _0}$ and $w_{n _{\alpha _F}}B_{\Gamma _{\alpha _F},\alpha _F}$. 

\vskip 0.5cm
\noindent
\textit{Explicit expression for star product}

\begin{figure}[t]
\begin{center}
\includegraphics[width=12cm,clip]{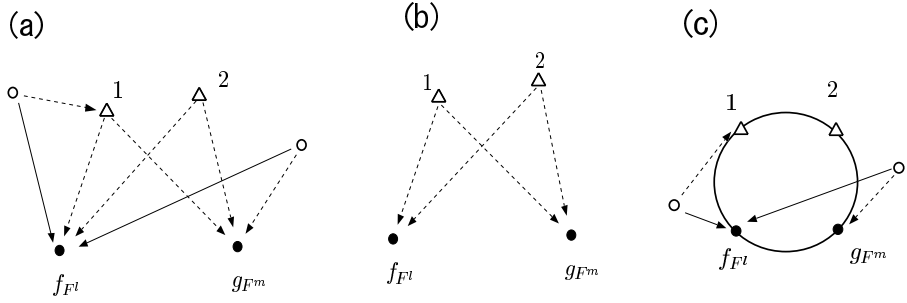}
\end{center}
\caption{
This figure shows the calculation method of the graph $(a)$, where the dotted arrow and real arrow represent the derivative with respect to $\pi $ and $X$, respectively, and the white circle and the white triangle represent $\alpha _0$ and $\alpha _F$, respectively. 
We rewrite the graph $(a)$ as the graph $(c)$ which is given by the cluster represented by the big circle and the operators into it, where the big circle represents the graph $(b)$. 
}\label{GD2}
\end{figure}

We will proceed to obtain the main result of this paper (Eq.~(\ref{result})) for the gauge-covariant star product. 
We treat $\alpha _F$, $f_{F^l}$ and $g_{F^m}$ as a cluster, and $\alpha _0$ as an operator acting on the cluster as shown in Fig.~\ref{GD2}. 
In this figure, the operator corresponding to the graph $(a)$ is calculated by two steps. 
First, the graph $(b)$ corresponding to $w_{n _{\alpha _F}}B_{\Gamma _{\alpha _F},\alpha _F}$ is treated as a cluster, and is given by Kontsevich's rules. 
Secondly, we calculate the operator $w_{n _{\alpha _0}}B_{\Gamma _{\alpha _0},\alpha _0}$ that acts on the cluster as in graph $(c)$, where the big open circle represents the graph $(b)$. 
Namely,
\begin{eqnarray}
(f\star g)_{F^r}|_{\genfrac{}{}{0pt}{}{f=f_{F^l}}{g=g_{F^m}}}=
\sum _{n_{\alpha _0},\Gamma _{\alpha _0}\in G^\prime }
\left (\frac{i\hbar }{2}\right )^nw_{\Gamma _{\alpha _0}}B_{\Gamma _{\alpha _0},\alpha _0}\cdot w_{\Gamma _{\alpha _F}}B_{\Gamma _{\alpha _F},\alpha _F}(f_{F^l},g_{F^m}),\nonumber\\
\end{eqnarray}
where $G^\prime $ is the set of $\Gamma _{\alpha _0}$. 

We consider the contribution of the cluster. 
Since edges of $\alpha _F(X)$ correspond to the derivative with respect to $\pi ^\mu $, $\alpha _F$-vertices do not connect with each other. 
In this case, the operator $w_{n _{\alpha _F}}B_{\Gamma _{\alpha _F},\alpha _F}$ is given by
\begin{eqnarray}
w_{n _{\alpha _F}}B_{\Gamma _{\alpha _F},\alpha _F}(f_{F^l},g_{F^m})=
\left (
\frac{2}{i\hbar }
\right )^{n_{\alpha _F}}
\frac{1}{n_{\alpha _F}!}
f_{F^l}(\frac{i\hbar }{2}\overleftarrow \partial _{\pi ^\mu }qF^{\mu \nu }\overrightarrow \partial _{\pi ^\nu })^{n_{\alpha _F}}
g_{F^m}. \nonumber\\ \label{B}
\end{eqnarray}

Using Eq.~(\ref{B}), $(f\star g)_{F^r}|_{\genfrac{}{}{0pt}{}{f=f_{F^l}}{g=g_{F^m}}}$ is given by 
\begin{eqnarray}
(f\star g)_{F^r}|_{\genfrac{}{}{0pt}{}{f=f_{F^l}}{g=g_{F^m}}}
&=&\sum _{n,\Gamma \in G_n}\left (\frac{i\hbar }{2}\right )^n
w_{n _{\alpha _0},n_{\alpha _F}}B_{\Gamma _{\alpha _0},\alpha _0}\cdot w_{n _{\alpha F}}B_{\Gamma _{\alpha _F},\alpha _F}(f_{F^l},g_{F^m})\nonumber\\
&=&\sum _{n_{\alpha _0}}\frac{(n_{\alpha _0}+n_{\alpha _F})!}{n_{\alpha _0}!n_{\alpha _F}!}
\left (\frac{i\hbar }{2}\right )^{n_{\alpha _0}+n_{\alpha _F}}
\frac{n_{\alpha _0}!n_{\alpha _F}!}{(n_{\alpha _0}+n_{\alpha _F})!}
\frac{\hat \alpha _0^{n_{\alpha _0}}}{n_{\alpha _0}!}w_{\Gamma _{\alpha F}}B_{\Gamma _{\alpha _F},\alpha _F}(f_{F^l},g_{F^m})
\nonumber\\
&=&\exp \left (\frac{i\hbar }{2}{\hat \alpha }_0\right )\left (\frac{1}{n_{\alpha _F}!}
f_{F^l}(\frac{i\hbar }{2}\overleftarrow \partial _{\pi ^\mu }qF ^{\mu \nu }\overrightarrow \partial _{\pi ^\nu })^{n_{\alpha _F}}
g_{F^m}\right ),\nonumber\\ \label{fg}
\end{eqnarray}
where $\hat \alpha _0$ is defined by 
\begin{eqnarray}
\hat \alpha _0(f_1,f_2,\cdots ):=\sum _{1\leq k_1<k_2<\infty }\sum _{i,j}\alpha _0^{ij}f_1\cdots (\partial _{x^i}f_{k_1})\cdots (\partial _{x^j}f_{k_2})\cdots .\nonumber\\ 
\end{eqnarray}
Thus the $r$th-order terms $(f\star g)_{F^r}$ is given by
\begin{eqnarray}
(f\star g)_{F^r}=\sum _{\genfrac{}{}{0pt}{}{l,m\in {\bf N}}{0\leq l+m\leq r}}
(f\star g)_{F^r}|_{\genfrac{}{}{0pt}{}{f=f_{F^l}}{g=g_{F^m}}}.\label{fg3.5} 
\end{eqnarray}
Substituting Eqs.~(\ref{fg}) and (\ref{fg3.5}) into Eqs.~(\ref{gauge-invDyson1}) and (\ref{gauge-invDyson2}), we obtain the gauge-covariant 
Dyson equations in the presence of  $F^{\mu \nu }$ as

\begin{eqnarray}
(f\star g)(X,\pi )=e^{\frac{i\hbar }{2}\hat \alpha _0}(f(X,\pi )e^{\frac{i\hbar }{2}\overleftarrow \partial _{\pi ^\mu }qF^{\mu \nu }\overrightarrow \partial _{\pi ^\nu }}g(X,\pi )).\label{result} 
\end{eqnarray}
When $f$, $g$ and $F$ are expanded by Fourier series, the operator $e^{\frac{i\hbar }{2}\hat \alpha _0}$ equals to the translation operator in the momentum 
space. 
We use this fact in the following section. 

\section{Linear response theory}\label{Linear}
As an application of the formalism developed above, we discuss in this section the linear response to the electromagnetic field $F^{\mu\nu}\propto e^{iQ_\mu X^\mu}$ starting from the Dyson equation. 
The linear components of Eq.~(\ref{gauge-invDyson1}) read 
\begin{eqnarray}
((\underline {\hat G}^{(0)-1}_0-\underline {\hat \Sigma })\star 
\underline {\hat G})_F(X,\pi )=0,\label{QBE1}\\
(\underline {\hat G}\star 
(\underline {\hat G}^{(0)-1}_0-\underline {\hat \Sigma }))_F(X,\pi )=0,\label{QBE2}
\end{eqnarray}
where $\underline {\hat G}^{(0)-1}_0(\pi )=\pi ^0-\hat H$(\mbox{\boldmath $\pi$}), $\hat H$ is the Hamiltonian of electrons and the linear order terms in $F_{\mu\nu}$ can be written as
\begin{eqnarray}
(f\star g)_F&:=&(f_F\star _0g_0)+(f_0\star _0g_F)+(f_0\star _Fg_0),\label{QBE3}
\end{eqnarray}
Each term is defined by 
\begin{eqnarray}
(f_F\star _0g_0)&:=&e^{\frac{i\hbar }{2}\hat \alpha _0}(f_F(X,\pi )g_0(X,\pi )),\label{product1}\\
(f_0\star _0g_F)&:=&e^{\frac{i\hbar }{2}\hat \alpha _0}(f_0(X,\pi )g_F(X,\pi )),\label{product2}\\
(f_0\star _Fg_0)&:=&e^{\frac{i\hbar }{2}\hat \alpha _0}
(f_0(X,\pi )(\frac{i\hbar }{2}\overleftarrow \partial _{\pi _\mu }qF^{\mu \nu }\overrightarrow \partial _{\pi _\nu })g_0(X,\pi )),\label{product3}
\end{eqnarray} 
where $f$ and $g$ represent $\underline {\hat G}$ or $({\underline {\hat G}^{(0)-1}_0}-\underline {\hat \Sigma })$. 
Hereafter, we assume that the equilibrium system is uniform, and the nonequilibrium one is steady, i.e., $f_F(X,\pi ),g_F(X,\pi )\propto e^{iQ_\mu X^\mu}$, $f_0(X,\pi )=f_0(\pi )$ and $g_0(X,\pi )=g_0(\pi )$. 
In this case, Eqs.~(\ref{product1}), (\ref{product2}) and (\ref{product3}) turn into 
\begin{eqnarray}
(f_F\star _0g_0)&=&f_F(X,\pi )g_0(\pi -Q/2),\\
(f_0\star _0g_F)&=&f_0(\pi +Q/2)g_F(X,\pi ),\\
(f_0\star _Fg_0)&=&\frac{i\hbar }{2}
f_0(\pi +Q/2)\overleftarrow \partial _{\pi _\mu }qF^{\mu \nu }\overrightarrow \partial _{\pi _\nu }g_0(\pi -Q/2),
\end{eqnarray} 
where $e^{a_\mu\partial _{\pi _\mu}}$ in $e^{\frac{i\hbar }{2}\hat \alpha _0}$ shifts the momentum $\pi _\mu$ to $\pi _\mu +a_\mu $. 
From Eq.~(\ref{QBE1}), the equation of the Green functions is given by
\begin{eqnarray}
\underline {\hat G}_F(X,\pi )&=&
\underline {\hat G}_0(\pi +Q/2)\underline {\hat \Sigma }_F(X,\pi )\underline {\hat G}_0(\pi -Q/2)\nonumber\\
&&-\underline {\hat G}_0(\pi +Q/2)
((\underline {\hat G}_0(\pi +Q/2)\overleftarrow \partial _{\pi _\mu }qF^{\mu \nu }\overrightarrow \partial _{\pi _\nu }\underline {\hat G}_0(\pi -Q/2)).\label{G1}\nonumber\\
\end{eqnarray}
We note that Eq.~(\ref{G1}) is rewritten by
\begin{eqnarray}
&&(\hat G^<_F-(n_{\rm Fermi}\star _0\hat G^A_F-\hat G^R_F\star _0n_{\rm Fermi}))
=\hat G^R_0\star _0(\hat \Sigma ^<_F-(n_{\rm Fermi}\star _0\hat \Sigma ^A_F-\hat \Sigma ^R_F\star _0n_{\rm Fermi} ))\star _0\hat G^A_0\nonumber\\
&&\ \ \ \ \ \ \ \ \ 
+\hat G^R_0\star _0n_{\rm Fermi}\star _0((\hat G^A_0)^{-1}\star _F\hat G^A_0)
-\hat G^R_0\star _0((\hat G^R_0)^{-1}\star _F\hat G^R_0)\star _0n_{\rm Fermi}\nonumber\\
&&\ \ \ \ \ \ \ \ \ 
-\hat G^R_0\star _0((\hat G^R_0)^{-1}\star _F\hat G^<_0)
+\hat G^R_0\star _0\hat \Sigma ^<_0\star _F\hat G^A_0,
\end{eqnarray}
where $n_{\rm Fermi}$ is the Fermi distribution function and $G^<_0=(G^A_0-G^R_0)n_{\rm Fermi}$. 
We have used the retarded and the advanced components of Eqs.~(\ref{QBE1}) and (\ref{QBE2}):
\begin{eqnarray}
&&(\hat G^{R,A}_0)^{-1}\star _0\hat G^{R,A}_F+(\hat G^{R,A}_F)^{-1}\star _0\hat G^{R,A}_0+(\hat G^{R,A}_0)^{-1}\star _F\hat G^{R,A}_0=0,\\
&&\hat G^{R,A}_F\star _0(\hat G^{R,A}_0)^{-1}+\hat G^{R,A}_0\star _0(\hat G^{R,A}_F)^{-1}+\hat G^{R,A}_0\star _F(\hat G^{R,A}_0)^{-1}=0.
\end{eqnarray} 
Therefore we obtain the following equations: 
\begin{eqnarray}
&&\hat G^{I}_F:=\hat G^<_F-(n_{\rm Fermi}\star _0\hat G^A_F-\hat G^R_F\star _0n_{\rm Fermi}),\\
&&\hat G^{II}_F:=n_{\rm Fermi}\star _0\hat G^A_F-\hat G^R_F\star _0n_{\rm Fermi},\\
&&\hat \Sigma ^{I}_F:=\hat \Sigma ^<_F-(n_{\rm Fermi}\star _0\hat \Sigma ^A_F-\hat \Sigma ^R_F\star _0n_{\rm Fermi}),\\
&&\hat \Sigma ^{II}_F:=n_{\rm Fermi}\star _0\hat \Sigma ^A_F-\hat \Sigma ^R_F\star _0n_{\rm Fermi},\\
&&\hat G^I_F=\hat G^R_0\star _0\hat \Sigma ^I_F\star _0\hat G^A_0\nonumber\\
&&\ \ \ \ \ \ \ \ \ 
+\hat G^R_0\star _0n_{\rm Fermi}\star _0((\hat G^A_0)^{-1}\star _F\hat G^A_0)
-\hat G^R_0\star _0((\hat G^R_0)^{-1}\star _F\hat G^R_0)\star _0n_{\rm Fermi}\nonumber\\
&&\ \ \ \ \ \ \ \ \ 
-\hat G^R_0\star _0((\hat G^R_0)^{-1}\star _F\hat G^<_0)
+\hat G^R_0\star _0\hat \Sigma ^<_0\star _F\hat G^A_0.\label{GI}
\end{eqnarray}
Eq.~(\ref{GI}) is the generalization of the St{\u r}eda formula~\cite{ptp1,streda}.  
\vskip 0.5cm

The work was supported by Grant-in-Aids under the Grant numbers 15104006,
16076205, and 17105002, and NAREGI Nanoscience Project from the Ministry of
Education, Culture, Sports, Science, and Technology.

\end{document}